\documentclass[a4 paper, 12 pt] {article}
\usepackage{amsmath}          % biblioteka z komendami matematycznymi
\usepackage{epsf}             % postscript (lamerskie)
\usepackage{amssymb}          % symbole mat
\usepackage{dsfont}           % liczba rzeczywista \mathds{R}
\usepackage{graphicx}         % wklejanie post
\usepackage{color}            % hmmm nie mam pojecia
\usepackage{rotate}           % tez ciekawa zagadka moze nie byc w

\newcommand{\opL}{{\bf L}}
\newcommand{\be}{\begin{equation}}
\newcommand{\ee}{\end{equation}}
\newcommand{\rev}[1]{(\ref{#1})}
\newcommand{\lab}[1] {\label{#1}}

\newcommand{\teta}{\tilde{\eta}}
\newcommand{\tf}{\tilde{T}}

\newcommand{\Umar}[1] {}

\begin{document}
\definecolor{myblue}{rgb}{0,0,0.8}

\title{\bf{Interaction Between Kink and Radiation in $\phi^4$
 Model}\thanks{Work performed under auspices of ESF Programme
 "Coslab"}}
\author{Tomasz Roma\'nczukiewicz\thanks{trom@th.if.uj.edu.pl}
       \\Institute of Physics,\\
       Jagiellonian University, Reymonta 4, Krakow, Poland}

\maketitle
\begin{abstract}
The radiation from oscillating kink in (1+1) dimensional
relativistic $\phi^4$  model is considered. Both analytical and
numerical approaches are presented and the comparison between
these methods is discussed. Acceleration of the kink in external
radiation is calculated and numerical results are also presented.
\end{abstract}

%%%%%%%%%%%%%%%%%%%%%%%%%%%%%%%%%%%%%%%%%%%%%%%%%%%%%%%%%%%%%%%%%%%%%%
%
\section{\bf{Introduction}}
%%%%%%%%%%%%%%%%%%%%%%%%%%%%%%%%%%%%%%%%%%%%%%%%%%%%%%%%%%%%%%%%%%%%%%
%
Kink is a simple example of topological defect. Topological
defects i.e. monopoles, vortices, domain walls play very important
role in modern physics starting from flux tubes in QCD \cite{QCD},
vortices in liquid helium and superconductors, domain walls in
magnetics, various defects in liquid crystals to strings in
cosmology \cite{Cosmic} (although recently it has been found
cosmological strings are not as important as they had appeared to
be \cite{Saka}). Although these objects were studied for many
years there are still many unanswered questions, mostly concerning
their dynamics. There are some open problems about their
interactions with other topological defects and external fields
\cite{Kisel}. Even dynamics of their internal degrees of freedom
is not fully understood and needs a lot of examinations
\cite{Arodz1}, \cite{Arodz2}. In most cases we built our knowledge
from experiments and numerical calculations. In theory, topological
defects are described by nonlinear partial differential equations.
Most of them one can solve exactly only in special cases, for
the rest of them one can apply only approximations.\\
In this paper we focus on perturbation method around a well known
analytical static solution namely the kink. The results will be
compared with numerical calculations. Very similar problems were
discussed in \cite{Manton}, \cite{Pelka} and \cite{Slus}.

Manton in \cite{Manton} considered problem of kink and antikink
creation and annihilation process and its relationship with the
vibrational mode. He presented theory concerning the behavior of
that mode and its decay. In this paper we develop methods
presented in \cite{Manton} and apply to different initial
conditions, more suitable for computer simulations. We verify
Manton's and our predictions via numerical calculations. In
\cite{Pelka} there are another methods for calculating the
radiation presented in a context of domain wall. The radiation
from squeezed kink is investigated in \cite{Slus}.

Usually when one wants to find out whether the system is stable,
one adds small perturbation and study its evolution. If the
perturbation does not grow in time it means the solution is
stable. Otherwise it is unstable. The problems of stability and
relaxation process is well examined for dissipative systems (for
example systems described by diffusion equations or wave equations
with damping). The model considered in the present paper is energy
conserving and therefore we can use stability only in a local
sense. The relaxation process must be based upon radiation of the
redundant energy into infinity. The static kink solution has the
lowest possible energy in its topological sector (Bogomolny
bound). This is a reason for us to treat kink as an attractor. The
system with localized perturbation will tend to our kink solution
in finite regions of the one-dimensional space.

In the sections 4 and 5  of our paper we apply methods presented in the
sections 2 and 3 to kink interacting with external
radiation. Because the kink is transparent to radiation in linear
order of approximation our usual intuition fails. For small
amplitudes, the kink instead of being pushed away by radiation
pressure is accelerating toward the source of radiation.

In literature there are considered mostly kinks interacting
with a constant external force \cite{Kisel} or a force oscillating in time
\cite{Sengupta} or \cite{Habib}.

\section{The model}
\begin{figure}
\begin{center}
\begin{picture}(400,200)
\put(10,0){\rotate[r]{\resizebox{8cm}{!}{\includegraphics{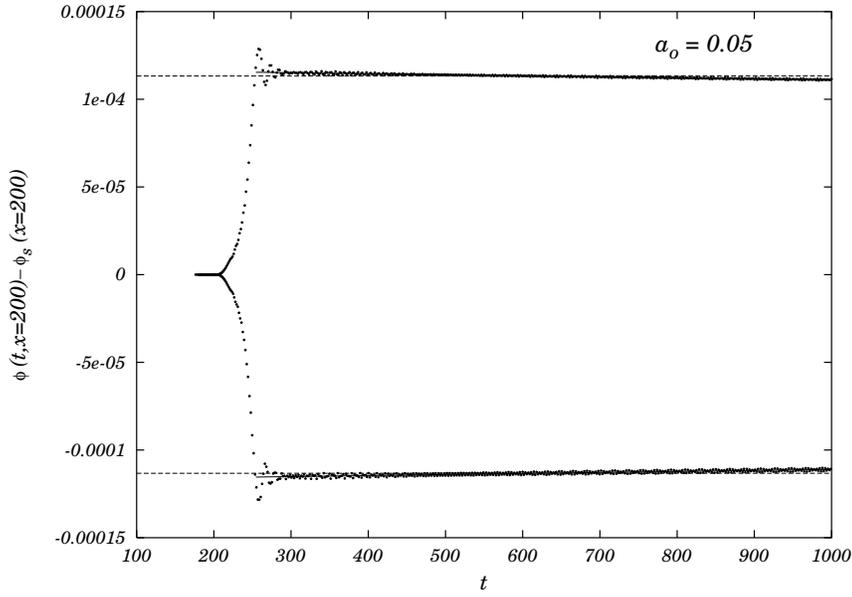}}}}
\end{picture}\\
\caption{Extrema of the outgoing radiation for $a_o=0.05$.} \label{f2}
\end{center}
\end{figure}

Let us consider one dimension real scalar field theory described by the
equation: \be
    \ddot{\phi}-\phi''+2\phi(\phi^2-1)=0. \lab{rI}
\ee
We use natural unit system ($c=1$).
There exist well known static solutions:
\be
   \phi_s(x) = \pm\tanh(x-x_o) \lab{rII}
\ee which will be referred to as kink (with $+$ sign) and antikink
(with $-$). Without loosing generality one can choose $x_o=0$ due
to translational invariance. Let us add a small perturbation to
the kink solution: \be
   \phi(t,x)=\phi_s(x)+\xi(t,x) \lab{rIII}.
\ee Eq. \rev{rI} can
be rewritten in the form:
\be
   \ddot{\xi}+\opL\xi+N(\xi)=0,
   \lab{rIV}
\ee
where linear operator $\opL$ has the form:
\be
   \opL=-\frac{\partial^2}{\partial
   x^2}+\left[4-6\left(1-\phi_s^2\right)\right],  \lab{rV}
\ee
and $N$ denotes the  part nonlinear in$ \xi$:
\be
   N(\xi) = 6\phi_s\xi^2+2\xi^3.
   \lab{rVI}
\ee In the first approximation we assume that $\xi$ is small
enough to neglect the term \rev{rVI}. We seek solutions of the
linearized eq. \rev{rIV} in the form $\xi(t,x)=e^{i\omega
t}\eta(x)$. We obtain the  eigenvalue problem (very similar to the
Schr\"odinger equation): \be \opL\eta=\omega^2\eta \lab{rVII}. \ee
The above equation has two solutions vanishing in infinity. One of
them for $\omega=0$
\be
  \eta_t(x)=\frac{1}{\cosh^2 x} \lab{rVIII}
\ee
is called translational zero--mode because it is responsible
for small translations of the kink: $\phi_s(x+\delta
x)=\phi_s(x)+\delta x\eta_t(x)+\mathcal{O}(\delta x^2)$. This mode
plays a very important role when one considers the evolution of a
system with an external force \cite{Kisel} or evolution of a
system itself in more dimensions (domain wall) \cite{Arodz1} or
interaction between two or more kinks. The other normalizable
solution is a vibrational mode with $\omega_d=\sqrt{3}$:
\be
  \eta_d(x)=\frac{\sinh x}{\cosh^2x}. \lab{rIX}
\ee
It is very
important that in linear approximation this mode oscillates
without loosing energy. It is a quasistationary solution because
only couplings in higher orders  re responsible for decaying of this solution
due to radiation. The energy is carried away by scattering modes
in continuous spectrum:
\be
 \eta_k(x)=e^{ikx}\left(3\tanh^2x-3ik\tanh x-1-k^2\right), \lab{rX}
\ee
where $k$ is a wave
vector: \be k^2=\omega^2-4. \ee We want to find the evolution of
the system when initially only the discrete oscillating mode is
excited. Let us construct a solution of the equation \rev{rIV}
substituting $\xi(t,x)=A_d(t)\eta_d(x)+\eta(t,x)$ and assuming
that $A_d$ and $\eta$ are small we obtain the equation for $\eta$
in $\mathcal{O}(A_d^2)$ order: \be
\ddot{\eta}+\opL\eta+6\phi_sA_d^2\eta_d^2=0. \lab{rXI} \ee This is
an inhomogeneous linear equation for $\eta$ with the source term
$T(t)g(x)$ where $g(x)=\eta^2_d(x)\phi_s(x)$ and $T(t)=6A_d^2(t)$. Notice that it is proportional to the square of vibrational mode. We can make a time Fourier transform of the
equation \rev{rXI} and obtain: \be
-\omega^2\teta(\omega,x)+\opL\teta(\omega,x)+\tf(\omega)g(x)=0.
\lab{rXII} \ee It is an  inhomogeneous linear equation for the spatial
part but since we know the solutions \rev{rX} of the homogeneous equation
we can easily construct appropriate Green's function in a
standard manner: \be G_k(x,y)=\begin{cases}
-\frac{1}{W}\eta_k(y)\eta_{-k}(x)&x<y\\
-\frac{1}{W}\eta_{-k}(y)\eta_{k}(x)&x>y
\end{cases}
\ee where $W=-2ik(k^2+1)(k^2+4)$ is Wronskian of eq.
\rev{rXII}. Therefore the solution has a form:
\be
  \teta(\omega,x)=-\frac{\tf(\omega)}{W}
  \left(\eta_{-k}(x)\int_{-\infty}^x\!\!dy\,\eta_k(y)g(y)+
  \eta_{k}(x)\int^{\infty}_x\!\!dy\, \eta_{-k}(y)g(y)\right).
  \lab{rXIII}
\ee
We can use an asymptotic form of $\eta_{\pm
k}(x)\approx(2-k^2\pm3ik)e^{\pm ikx}$ for large $x$ and obtain
(the source is well localized around 0):
\be
  \teta(\omega,x)=-\frac{\tf(\omega)}{2ik(2-k^2-3ik)}e^{(-ikx)}
  \int_{-\infty}^{\infty}\!\!\!dy\, \eta_k(y)g(y).
\ee
After calculating this integral analytically we get:
\be
  \teta(\omega,x)=-\frac{\pi k(k^2+4)(k^2-2)\tf(\omega)}
  {16\sinh\left(\frac{\pi k}{2}\right)(2-k^2-3ik)}\exp(-ikx)
  \lab{rXIV}
\ee
Following Manton \cite{Manton} we consider an
oscillating mode $A_d(t)=a\cos(\omega_d t)$. Square of the
amplitude is given by
$A_d^2=\frac{1}{2}a^2\left(\cos(2\omega_dt)+1\right)$. Since
constant solution carries no energy we take into account only the
time dependent part: $T(t)=3a^2e^{2\omega_dt}$. Fourier
transform has a form: $\tf(\omega)=-6\pi
a^2\delta(\omega-\omega_o)$, where $\omega_o=2\omega_d$.
Substituting the source term into \rev{rXIV} and calculating
inverted Fourier transform we obtain:
\be
  \eta(t,x)=\frac{\pi
  k_o(k_o^2+4)(k_o^2-2)} {32\sinh\left(\frac{\pi k_o}{2}\right)
  (2-k_o^2-3ik_o)} a^2\exp\left[i(\omega_ot-k_ox)\right].\lab{rXV}
\ee
Since we are only interested in real part of the above
equation the radiation has a form:
\be
  \eta(t,x)=Qa^2\cos\left(\omega_ot-k_ox+\delta \right),\lab{rXVI}
\ee
where
$$
k_o^2=\omega_o^2-4,
$$
\be
  Q=\frac{\pi k_o(k_o^2-2)} {32\sinh\left(\frac{\pi
  k_o}{2}\right) } \sqrt{\frac{k_o^2+4}{k_o^2+1}}=0.0453,
\ee
and
$\delta$ is a phase which we are not interested in. The same
radiation goes also to $-\infty$. We have made numerical simulations
for various $a_o$, where $a_o$ is initial amplitude of the
oscillating mode (in the following section we show that the amplitude decreases due to the radiation). In large distance from the kink we
measure the outgoing field. The difference from the kink solution is
sketched in figures \ref{f2} and \ref{f1} for $a_o=0.05$ and $0.4$
respectively (for the clarity we have plotted only the local extrema
of the field instead of whole oscillations). The amplitude
predicted by eq. \rev{rXVI} seem to fit very well in the first figure.
On the second Figure one can also see the damping which will be
explained in the next section.

As we can see from eq. \rev{rXVI} the outgoing radiation has only
one single frequency which is twice the frequency of oscillating
mode. This is true only in approximation, because the
coupling in cubic term in \rev{rVI} will result in frequency
$3\omega_d$. If one wants to test this procedure numerically one
has to be very careful. The first idea is to pose initial
conditions in the form $\phi(0,x)=\phi_s(x)+a\eta_d(x)$ and
$\dot{\phi}(0,x)=0$. That means there is no radiation in $t=0$, so
there can be no source for $t<0$. Source must have a form
$T(t)=3a^2\Theta(t)e^{i\omega_ot}$, where $\Theta(x)$ is
a Heaviside function. Fourier transform:
\be
  \tf(\omega)=3a^2\left(-\pi\delta(\omega-\omega_o)+
  P\frac{i}{\omega-\omega_o}\right).
\ee
The field we calculate is equal to
\be
\eta(t,x)=\frac{a^2}{2\pi}\text{P.V.}\int_{-\infty}^\infty\!\!d\omega
  \; \frac{\pi k(k^2+4)(k^2-2)\exp(i(\omega t-kx))}
  {32\sinh\left(\frac{\pi
  k}{2}\right)(2-k^2-3ik)}\left(-\pi\delta(\omega-\omega_o)+
  \frac{i}{\omega-\omega_o}\right).\lab{rXVII}
\ee
Let us consider the second part of the integral in \rev{rXVII}
containing factor $\frac{i}{\omega-\omega_o}$:

\be
  I(t,x)=i\int_{-\infty}^{\infty}\!\!\!\! d\omega\,
  F(\omega)\frac{e^{i\omega t}}{\omega-\omega_o}.
\ee If $k$ is a real number we have oscillations of the integrand
along $x$ axis (because of the $\exp(-ikx)$ term), but
when $k$ is imaginary we have exponential decay, therefore for
large $x$ we need to take into account only $\omega$ for which $k$
is real, that is $\omega>\Omega$, where $\Omega=2$. We can rewrite
the above integral in a form: \be
  I(t, x)=ie^{i\omega_ot}\int_{\Omega}^\infty
  \!\!\!d\omega\,F(\omega)\frac{e^{i(\omega-\omega_o)t}}
  {\omega-\omega_o}
\ee
After introducing $\nu=(\omega-\omega_o)t$ we obtain:
\be
  I(t,x)=
  ie^{i\omega_ot}\int_{(\Omega-\omega_o)t}^\infty
  \!\!\!\!\!\!d\nu\,F\left(\frac{\nu}{t}+\omega_o\right)
  \frac{e^{i\nu}}{\nu}
\ee
For large times we can expand $F$ in $\frac{\nu}{t}$:
\be
  I(t,x)=
  ie^{i\omega_ot}\int_{(\Omega-\omega_o)t}^\infty
  \!\!\!\!\!\!d\nu\,F\left(\omega_o\right)
  \frac{e^{i\nu}}{\nu}+F'(\omega_o)\frac{e^{i\nu}}{t}+\cdots
\ee
Because $\Omega-\omega_o<0$ and $t$ tends to infinity the first term
is equal to:
\be
  \lim_{t\rightarrow\infty}\int_{(\Omega-\omega_o)t}^\infty
  \!\!\!\!\!\!d\nu\,F\left(\omega_o\right)
  \frac{e^{i\nu}}{\nu}=F(\omega_o)\int_{-\infty}^{\infty}
  \!\!\!d\nu \frac{e^{i\nu}}{\nu}=i\pi F(\omega_o)
\ee The second term we can integrate immediately after adding a
term $-\epsilon\nu$ to the exponent and obtain: \be
  I(t,x) = -e^{i\omega_ot}\left[\pi
  F(\omega_o)+\frac{e^{i(\Omega-\omega_o)t}}{t}F'(\omega_o)\right].
\ee
But $F(\omega)=e^{-ikx}f(\omega)$ therefore
$$
  F'(\omega_o)=\left(-ixf(\omega_o)\left.\frac{dk}{d\omega}\right|
  _{\omega=\omega_o} +f'(\omega_o)\right)e^{-ik_ox}.
$$
For large $x$ we can neglect the second term:
\be
  I(t,x)=-e^{i\omega_ot}\left[\pi
  -ix\frac{\omega_o}{k_o}\frac{e^{i(\Omega-\omega_o)t}}{t}
  \right]F(\omega_o).
\ee
Finally the radiating part in our approximation has a form
\be
  \eta(t,x)=Qa^2
  \left[1-i\frac{\omega_o}{k_o}
  \frac{e^{i(\Omega-\omega_o)t}}{2\pi t/x}\right]
  \exp\bigl[i(\omega_ot-k_ox+\delta)\bigr].
\ee Apart from oscillation with frequency $\omega_o$ we have some
decaying modulations with frequency $\omega_o-\Omega\approx1.46$.
Speed of the decay is determined by the point of observation $x$.
The larger $x$ the slower decay. Figure 3 shows the Fourier's
transform of the numerically computed field in $x=30$. As one can
see there are three easily visible peaks. The largest one is at
$\omega=\omega_o=2\omega_d$. The next one is at another harmonic
frequency $3\omega_d$. There is also one peak for
$\omega=1.15\omega_d=2=\Omega$ which is responsible for decaying
modulations of the field. On the next Figure we have plotted
envelope of oscillations evaluated from eq. \rev{rXVII}. On Figure
5 one can see the behavior of the envelope for large times which
seems to be well approximated by eq. (30). The measured frequency
is really $\omega_o-\Omega\approx1.42$. We also present the growth
of the field for small times  on figures 6 and 7. On the next
pictures we have sketched numerically evaluated field (from eq.
\rev{rXVII}) and, for comparison, numerically calculated solution
of the full partial equation (1). Because of the numerical errors
there are only the largest modulations visible.

\section{Backreaction}
In the above approximation we have found the field radiated to the
infinity. We assumed that amplitude of the oscillating mode does
not change in time. This is not true if the system is isolated. Radiation carries away energy and that leads to a decrease of the amplitude since the system
is energy conserving. Let us calculate the rate at which the energy is being
lost during radiation. For scalar field the energy escaping from a
segment $(x_1, x_2)$ is equal to
$\frac{dE}{dt}=\bigl.\dot{\phi}\phi'\bigr|_{x_1}-\bigl.\dot{\phi}\phi'
\bigr|_{x_2}$. After taking
$\phi(t,x)=\phi_s(x)+a\eta_d(x)\cos(\omega_d t)+Qa^2\cos(\omega_o
t-k_ox+\delta)$ and averaging over a period we obtain the energy
change inside large segment: \be
\frac{dE}{dt}=-Q^2k_o\omega_oa^4=-0.020a^4 \ee and since energy of
the vibrational mode is $E=a^2$ we find a differential equation
for the amplitude of that mode: \be
\frac{da^2}{dt}=-Q^2k_o\omega_oa^4. \ee We can easily find a
solution with an initial condition $a(t=0)=a_o$: \be
a(t)=\frac{a_o}{\sqrt{1+Q^2k_o\omega_o a_o^2t}}. \lab{rXVIII} \ee
From the above we obtain time dependence of radiation amplitude:
\be A(t)=\frac{1}{C(x,a_o)+Qk_o\omega_ot},\lab{r1XVIII} \ee where
\be C(x,a_o)=\frac{1}{Qa_o^2}-Qk_o\omega_o t_o(x)\lab{RXVIII} \ee
is constant in time. $t_o(x)$ is the time it takes for the field
to travel from origin to point $x$ where it is observed.
% added to time from
%singularity which has of course no physical meaning because for
%large fields our small field approximation breaks down.

\begin{figure}
\begin{center}
\begin{picture}(400,200)
\put(10,0){\rotate[r]{\resizebox{8cm}{!}{\includegraphics{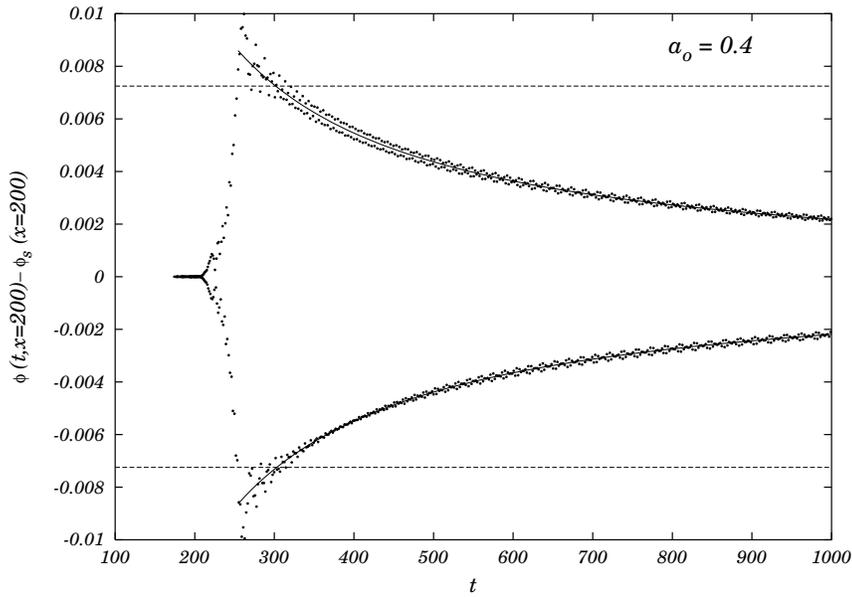}}}}
\end{picture}\\
\caption{Extrema of the decaying radiation for $a_o=0.4$ with fitted
functions \rev{r1XVIII}.}\label{f1}
\end{center}
\end{figure}

This very precise prediction we can compare with purely numerical
results. Maxima and minima of the field with initial $a_o=0.4$
measured for $x=200$ are shown on Figure \ref{f1}. There are also
two hyperbolic function fitted to these data. The same procedure
was applied for other amplitudes. The factor that stands besides
$t$ in eq. \rev{r1XVIII} in the theory is equal to 0.444. Numerical
results show that its average value is 0.449 which is not
significantly different. The value does not change much (within
2\%) for different $a_o$ (0.05$\div 0.7$).

In Figure \ref{fig2} we have sketched fitted $C$ versus $a_o$. As
one can see the fitted function in the form \rev{RXVIII} is a very
good approximation of the results. We find $Q=0.04607$. In the
theory above we have 0.0453 but it is still within numerical
error. The last parameter in numerics is equal to $t_o=336$. The
group velocity of the field is equal to
$v=\left.\frac{\partial\omega}{\partial
k}\right|_{k=k_o}=\frac{k_o}{\omega_o}\approx0.816$ and hence the
distance $x=200$ than theoretical value of $t_{o,th}=245$. In fact
we actually can observe that the radiation cames to $x=200$ in
this time (figure \ref{f2} and \ref{f1}). If one plots the fitted
function and the theoretical one on one graph one can find they
are almost  indistinguishable, especially for small $a_o$ where
our theory is the most accurate.

%Actually we do not have a good prediction for its value in theory.
%One can think of it as a time needed for the field to travel from
%its origin to the point where it is measured. But the velocity of
%the field is equal to $\frac{k_o}{\omega_o}\approx 0.816$ and the
%distance is 200, so $t_o$ should be equal to 245. In fact we
%actually can observe that the radiation cames to $x=200$ in this
%time (figure \ref{f2} and \ref{f1}). So probably $t_o$ is not
%connected with travel time in that simple way. Maybe different
%initial conditions would give some better results or more
%sophisticated analysis of the eq. \rev{rXVII}.

\section{Anharmonic corrections}
As we saw in previous section, the field radiated from oscillating
kink is of order $\mathcal{O}(A_d^2)$ (eq. \rev{rXVI},
$A_d=a\cos(\omega_o t)$). The oscillating mode reaction to scattering
modes will be of order $\mathcal{O}(A_d^3)$ (in eq. \rev{rXVIII}
$a(t)\approx a_o-\frac{1}{2}a_o^3Q^2k_o\omega_ot$). But there is still
nontrivial behavior of amplitude in the order $\mathcal{O}(A_d^2)$.
Let us consider only the discrete mode. Substituting
$\xi(t,x)=A_d(t)\eta_d(x)$ in eq. \rev{rIV} in order
$\mathcal{O}(A_d^2)$ we obtain the following equation:
\be
   \left(\ddot{A_d}+3A_d\right)\eta_d+6A^2_d\phi_s\eta^2_d=0.
\ee
We are interested only in evolution of $A_d(t)$.We can rewrite the
source term in a form $\phi_s\eta_d^2=\alpha\eta_d+\eta_{\perp}$. To
get rid of the orthogonal term $\eta_\perp$ we project this
equation onto $\eta_d$. In order to do that we integrate both sides of
the above equation with $\eta_d$ and we obtain an ordinary
differential second order equation (neglecting the perpendicular
part):

\be
  \ddot{A_d}+3A_d+\frac{9\pi}{16}A_d^2=0.
\ee

We solve this using standard series method:
$A_d=A^{(1)}+A^{(2)}+A^{(3)}+\cdots$. We take initial conditions:
$\dot{A}(0)=0,\;A(0)=a$. In order $\mathcal{O}(A^{(1)})$ we obtain

\be \ddot{A}^{(1)} +3A^{(1)}=0,\ee and the solution is
$A^{(1)}=a\cos\sqrt{3}t$. In second order we have:

\be
\ddot{A}^{(2)}+3A^{(2)}+\frac{9\pi}{16}a^2\cos^2\sqrt{3}t=0.\ee
The whole solution has a form:

\be
A_d(t)=-\frac{3\pi}{32}a^2+\left(1+\frac{a\pi}{16}\right)a\cos(\sqrt{3
}t)+
\frac{\pi}{32}a^2\cos(2\sqrt{3}t). \ee

This function's maxima are equal to $A_{max}=a$ and minima
$A_{min}=-a-\frac{3\pi}{16}a^2$.
Figure \ref{anh} shows the oscillating mode evolution calculated
numerically for $a_o=0.2$. The field was measured in
$x=x_{max}={\rm ar}\!\cosh \sqrt{2}\approx 0.88$, where the function
$\eta_d$ has maximum equal $\frac{1}{2}$. The dashed lines represent
$\frac{1}{2}A_{max}$ and $\frac{1}{2}A_{min}$ calculated for that
amplitude. As one can see our prediction works very well.\\
Because the radiation depends on the square of the oscillating
mode one can expect that minima are on more or less the same level
while the maxima are changing (one is higher and the next one is
lower and the following one is higher again). This effect is
actually seen on Figure \ref{f1}.
\newpage

\section{Kink in external radiation}
Let us now consider quite opposite process when kink is exposed to
external radiation coming from $+\infty$:  $ae^{i\omega_q
t}\eta_q(x)$. Let us consider small amplitude limit. In the
linear approximation there is no reflection in the
potential of the soliton and therefore one cannot expect the
system will behave as other similar systems in other branches of
physics (there is no analogy to  a particle exposed to an
electromagnetic wave). In the first order the solution has a form
\rev{rXIII}, where $\tf(\omega)=-3\pi a^2\delta(\omega-2\omega_q)$
and $g(x)=\eta^2_q(x)\phi_s(x)$. The correction oscillates with
frequency $2\omega_q$. For $x\longrightarrow\pm\infty$ we can
approximate the solution to form:
\be
\begin{aligned}
 &\teta(\omega,x\rightarrow\infty)=-\frac{\tf(\omega)}{W}\left(
 \eta_{-k}(x)\int^\infty_{-\infty}
 \!\!\!dy\,\eta_k(y)\eta_q^2(y)\phi_s(y)\right.\\
&\left.-\eta_{-k}(x)\int^\infty_{x}
 \!\!\!dy\,\eta_k(y)\eta_q^2(y)\phi_s(y)+\eta_{k}(x)\int^\infty_{x}
 \!\!\!dy\,\eta_{-k}(y)\eta_q^2(y)\phi_s(y)\right)
\end{aligned}
\ee
and
\be
\begin{aligned}
 &\teta(\omega,x\rightarrow-\infty)=-\frac{\tf(\omega)}{W}\left(
 \eta_{k}(x)\int^\infty_{-\infty}
 \!\!\!dy\,\eta_{-k}(y)\eta_q^2(y)\phi_s(y)\right.\\
&\left.-\eta_{k}(x)\int_{-\infty}^{x}
 \!\!\!dy\,\eta_{-k}(y)\eta_q^2(y)\phi_s(y)+\eta_{-k}(x)\int_{-\infty}^{x}
 \!\!\!dy\,\eta_{k}(y)\eta_q^2(y)\phi_s(y)\right)
\end{aligned}
\ee
%\be
%  \teta(\omega,x\rightarrow-\infty)=
%  -\frac{6\tf(\omega)}{W}\eta_{k}(x)\int^\infty_{-\infty}
%  \!\!\!dy\,\eta_{-k}(y)\eta_q^2(y)\phi_s(y)
%\ee
where $k=k(q)=\sqrt{(2\omega_q)^2-4}$. In order to calculate the
integrands with one finite but large limit we can use once again
the asymptotic form of $\eta_k(y)$. Integrands with both infinite
limits can be calculated analytically via residua and finally we
can write a simply expression for asymptotic behavior of the first
correction: \be
 \eta_{+\infty}(t,x)=b_+(\omega_q)\cos(\omega_kt-kx+\delta_1)+c(\omega_q)
 \cos(\omega_kt+2qt+\delta_2)
\ee
and
\be
 \eta_{-\infty}(t,x)=b_-(\omega_q)\cos(\omega_kt+kx-\delta_1)-c(\omega_q)
 \cos(\omega_kt+2qt-\delta_2).
\ee
where
\be
  \hspace*{-1cm}b_\pm(\omega_q)=3\pi
  a^2\frac{-60480+82224\omega_q^2-31620\omega_q^4+3707\omega_q^6\pm
  qk(15120-11115\omega_q^2+1851\omega_q^4)}
  {10\sinh\left(\displaystyle\frac{k\pm2q}{2}\pi\right)(4\omega_q^2-3)
  \omega_q^2k},
\ee \be c(\omega_q)=\frac{3}{2k}(q^2+1)\omega_q^2 \ee and
$\delta_{1,2}$ are phases. The reflected part $b_+$ for small
frequencies is more or less the same order as $b_-$, but for large
frequencies decays exponentially
$b_+(\omega_q)\sim\frac{22227}{80}\pi\omega_q
a^2e^{-2\pi\omega_q}$. The transition part
$b_-\sim\frac{1}{8}\omega^2_qa^2$. $c$ is an amplitude of a wave
coming from $\infty$ and going to $-\infty$. This wave has exactly
the same form at both sides and hence passes no energy nor
momentum to the kink as one can see in details in the following
calculations.

For small $x$ we do not know an analytical solution. The solution
(which for small $x$ can differ from $\eta_{\pm k}$) oscillates
around the kink solution with a frequency of the source wave
$2\omega_q$. Because of the energy gathered in these oscillations
kink as seen from a distance  gains some extra mass. In following
calculations we will be refering to that effect mass as $M^*$. The
bare mass of the soliton equals $M=\frac{4}{3}$.
%First correction to the soliton mass is of order $\omega_q^2a^2$.
As mentioned in a section 3 the change of energy  inside a segment
$(x_1,x_2)$ equals $\frac{dE}{dt} =
\bigl.\dot{\phi}\phi'\bigr|^{x_2}_{x_1}$. In our case we have from
the right side
$\phi(x)=\phi_s(x)+A\cos(\omega_qt+qx)+b_+\cos(\omega_kt-kx)+c\cos(\omega_kt+2qt)$. The
energy flowing into the segment averaged over a period is
$$\frac{dE_{r}}{dt}=\frac{1}{2}q\omega_qA^2-\frac{1}{2}k\omega_kb_+^2+
q\omega_kc^2.$$
From the left side the field is
$\phi(x)=\phi_s(x)+B\cos(\omega_qt+qx)+b_-\cos\left(2\omega_qt+kx\right)-
c\cos(2\omega_kt+2qt)$.
The energy is equal than:
$$\frac{dE_{l}}{dt}=-\frac{1}{2}q\omega_qB^2-\frac{1}{2}k\omega_kb_-^2
-q\omega_kc^2.$$
The energy conservation takes a form: \be
\frac{1}{2}q\omega_qA^2-\frac{1}{2}q\omega_qB^2-\frac{1}{2}k\omega_k
\left(b_-^2+b_+^2\right)= M^*\frac{d\gamma}{dt}. \ee Where
$M^*\gamma=\frac{M^*}{\sqrt{1-v^2}}$ is an energy of the soliton.
Apart from $M^*$ which is probably different from $M$ we have two
unknown  kinetic variables $B$ and $\dot{\gamma}$. The second one
is responsible for the acceleration of the kink due to interaction
with the radiation. We need one more equation. We can use the
momentum conservation law:
$$\frac{dP_r}{dt}=\frac{1}{2}\omega_q^2A^2+\frac{1}{2}\omega_k^2b_+^2+\frac{1}{2}\omega_k^2c^2$$
and from the left side
$$\frac{dP_l}{dt}=-\frac{1}{2}\omega_q^2B^2-\frac{1}{2}\omega_k^2b_-^2
-\frac{1}{2}\omega_k^2c^2$$
and hence the second equation
\be
  \frac{1}{2}\omega_q^2A^2-\frac{1}{2}\omega_q^2B^2-\frac{1}{2}\omega_k^2
 \left(b_-^2-b_+^2\right) = M^*\frac{d}{dt}(v\gamma)=F^*
\ee

%We can rewrite the rhs of the equations using
%$\dot{\gamma}=-\gamma^3v\dot{v}$ and
%$(v\gamma)^\cdot=\gamma(1-v^2\gamma^2)\dot{v}$.
These equations are correct only for $v=0$ because the moving kink
experiences the Doppler's effect and there should be corrections
in $\omega_{q,k}$, $k$ and $q$. But substituting $v=0$ to above
equations we obtain the solutions very easily: \be
\begin{cases}
\displaystyle B^2&=A^2-\displaystyle 2\frac{k}{q}\left(b_-^2+b_+^2\right)\\
\\
\displaystyle F^*&=\displaystyle \frac{\omega_q^2}{q}
\left((k-2q)b_-^2+(k+2q)b_+^2\right)
\end{cases}
\ee We used the relation $\omega_k=2\omega_q$. The factor before
$b_-$ is
$$k-2q=2\left(\sqrt{\omega_q^2-1}-\sqrt{\omega_q^2-4}\right)$$
is always positive and therefore the soliton accelerates in the direction from which radiation comes. The radiation presure is negative! The force desribed by above equation is shown on Figure \ref{force}.\\
Because $b_+(\omega_q)$ is usually very small, for certain $\omega_q$  when
$b_-(\omega_q)=0$ the acceleration derived in this order of
approximation vanishes and nonlinearities of higher orders  play
the crucial role. Numerical results show the soliton will be pushed by the radiation.
The conclusion is that the soliton in most cases will stem the current.
If the soliton reaches the speed
when the frequency of the source wave due to the Doppler's effect
$\omega'=\gamma\left(\omega_q+v\sqrt{\omega_q^2-4}\right)$ is the
frequency for which acceleration vanishes the soliton moves with
more or less constant velocity. Figure \ref{wstecz1} shows the paths
of zeros of the field (kinks) for frequencies around the minimum
of the pulling force.
%For the most cases kink accelerates in positive direction.
If the source wave has a frequency very close
to the minimum of the force the kink's acceleration is small and
the soliton moves with the current.

If the source wave amplitude is large the above approximation
fails. There exists certain critical amplitude, depending on a frequency, for which the soliton is pushed away by the incoming wave (Figure \ref{wstecz2}).
% If one finds the velocity and
%calculates the observed frequency one finds that the above
%argumentation is justified.

\newpage
%%%%%%%%%%%%%%
\section{\bf{Conclusions}}
In this paper we have investigated the behavior of excited oscillating
mode and its decay using both analytical and numerical methods. We
have proved conformity between these two methods. We have studied
the interaction of the kink with external radiation field. We have found that in most cases the kink will be pulled by the radiation.
We have explained the peculiar behavior of this system.
Finally we presented numerical results for large amplitudes of the
source wave when the kink is pushed away by the wave.

%%%%%%%%%%%%%%

%\begin{center}
%\begin{picture}(400,250)
%\psfrag{"A=0.8bf.dat"}{}
%\put(10,0){\rotate[r]{\resizebox{9cm}{!}{\includegraphics{ft1.ps}}}}
%\put(200,-5){\small $\omega$}
%\put(8,100){\hspace*{-5mm}\rotate[l]{$|\tilde{\phi}(\omega)|$}}
%\end{picture}\\

\newpage

\begin{figure}
\begin{center}
\begin{picture}(400,200)
\put(10,0){\rotate[r]{\resizebox{8cm}{!}{\includegraphics{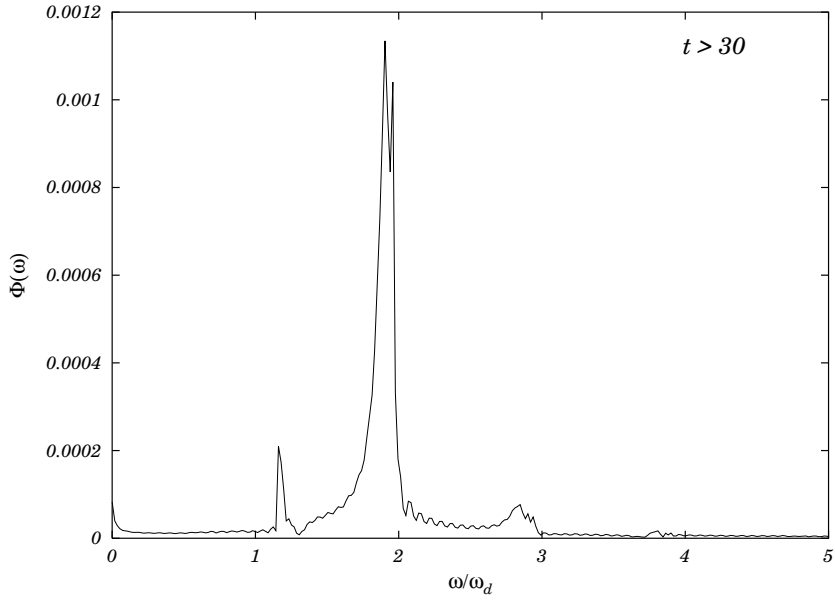}}}}
\end{picture}\\
\caption{Fourier transform of the radiation ($a_o=0.8$).}\label{fur1}
\end{center} \end{figure}

\begin{figure}
\begin{center}
\begin{picture}(400,200)
\put(10,0){\rotate[r]{\resizebox{8cm}{!}{\includegraphics{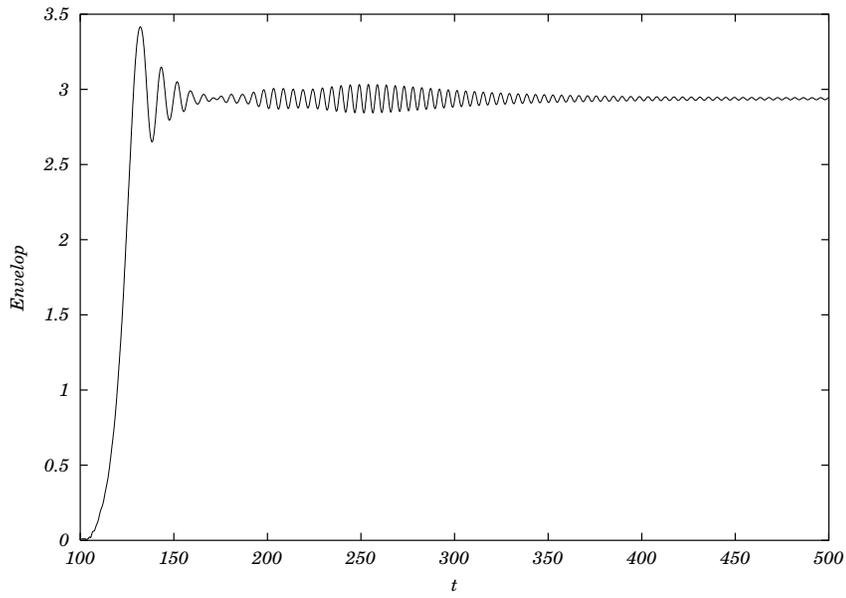}}}}
\end{picture}\\
\caption{Envelope of the radiation} \label{en1}
\end{center}
\end{figure}

\begin{figure}
\begin{center}
\begin{picture}(400,200)
\put(10,0){\rotate[r]{\resizebox{8cm}{!}{\includegraphics{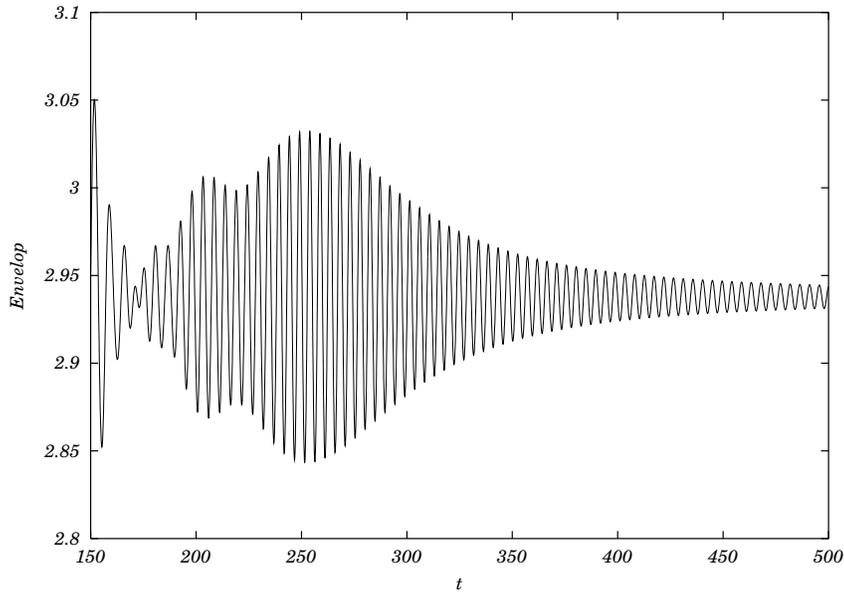}}}}
\end{picture}\\
\caption{Envelope for large times} \label{en2}
\end{center}
\end{figure}

\begin{figure}
\begin{center}
\begin{picture}(400,200)
\put(10,0){\rotate[r]{\resizebox{8cm}{!}{\includegraphics{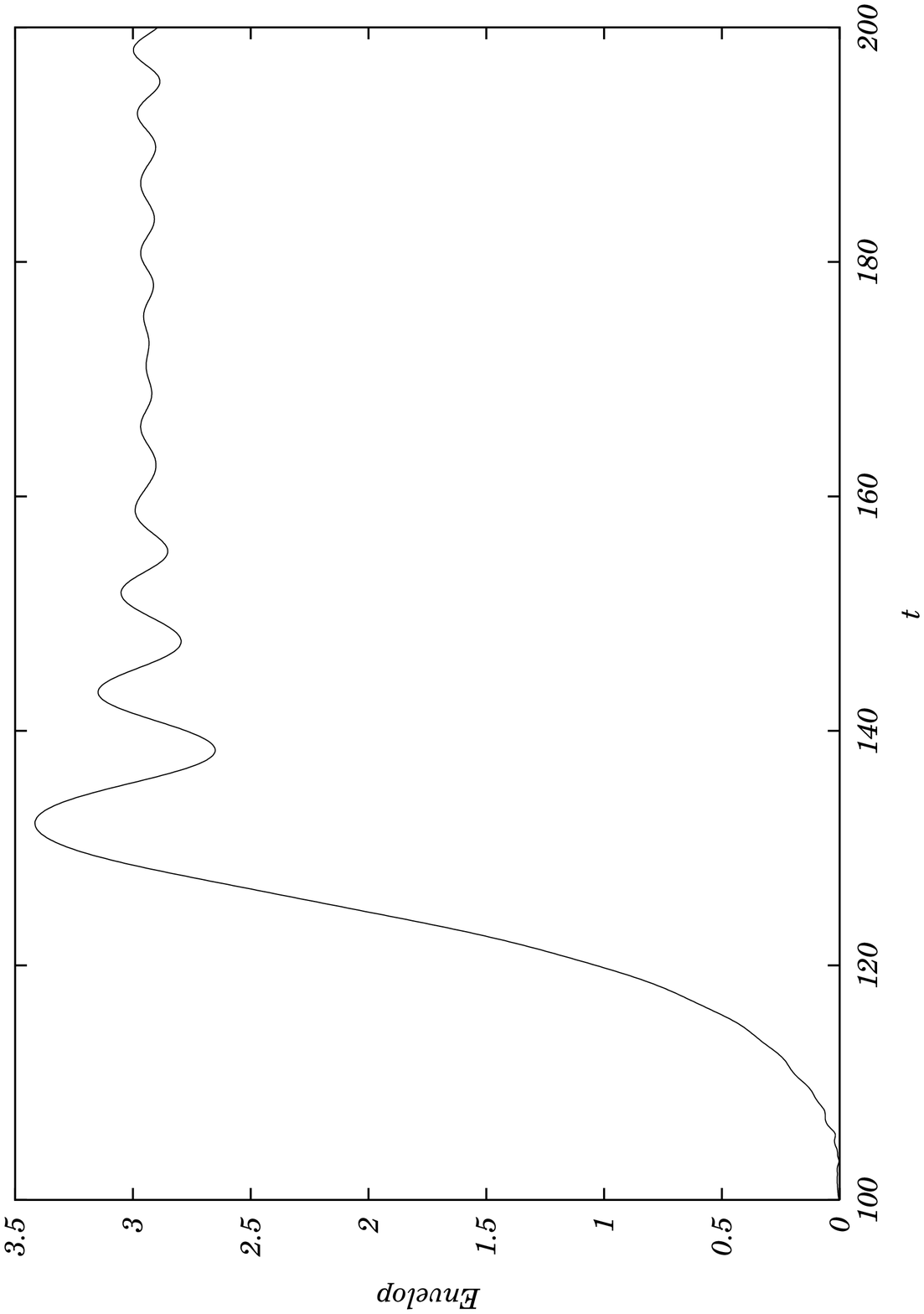}}}}
\end{picture}\\
\caption{Envelope for small times}\label{en3}
\end{center}
\end{figure}

\begin{figure}
\begin{center}
\begin{picture}(400,200)
\put(10,0){\rotate[r]{\resizebox{8cm}{!}{\includegraphics{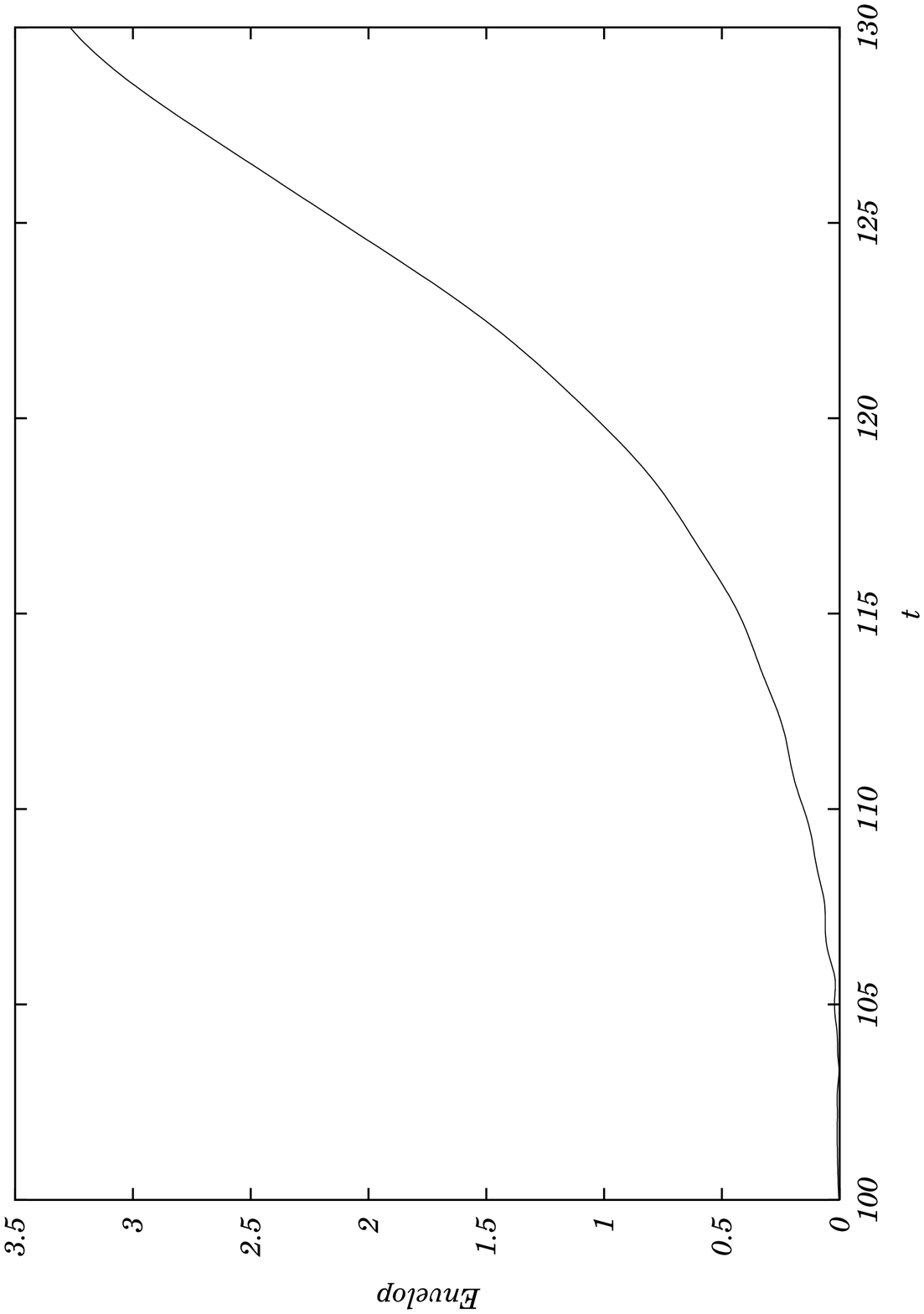}}}}
\end{picture}\\
\caption{Envelope for very small times}\label{en4}
\end{center}
\end{figure}

\begin{figure}
\begin{center}
\begin{picture}(400,200)
\put(10,0){\rotate[r]{\resizebox{8cm}{!}{\includegraphics{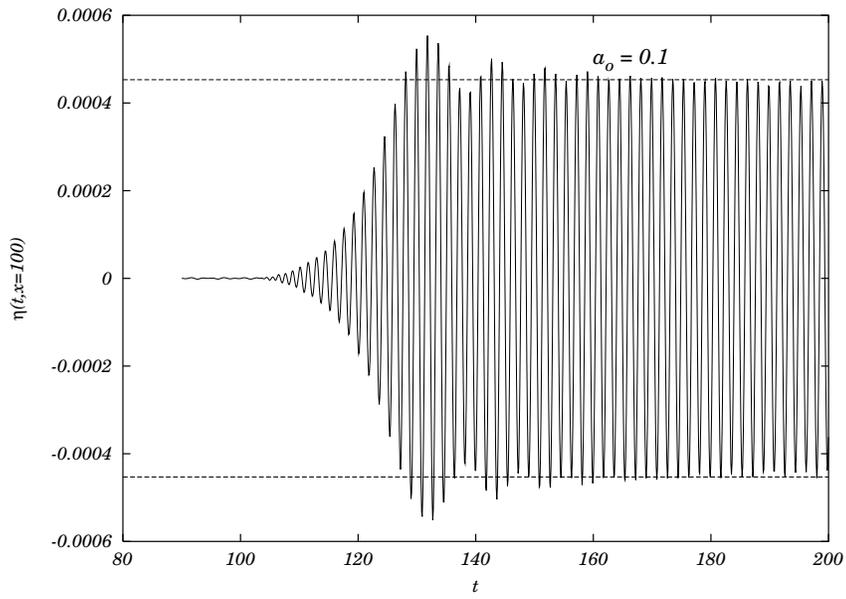}}}}
\end{picture}\\
\caption{Radiation calculated from integral (7)}\label{calk1}
\end{center}
\end{figure}
\begin{figure}

\begin{center}
\begin{picture}(400,200)
\put(10,0){\rotate[r]{\resizebox{8cm}{!}{\includegraphics{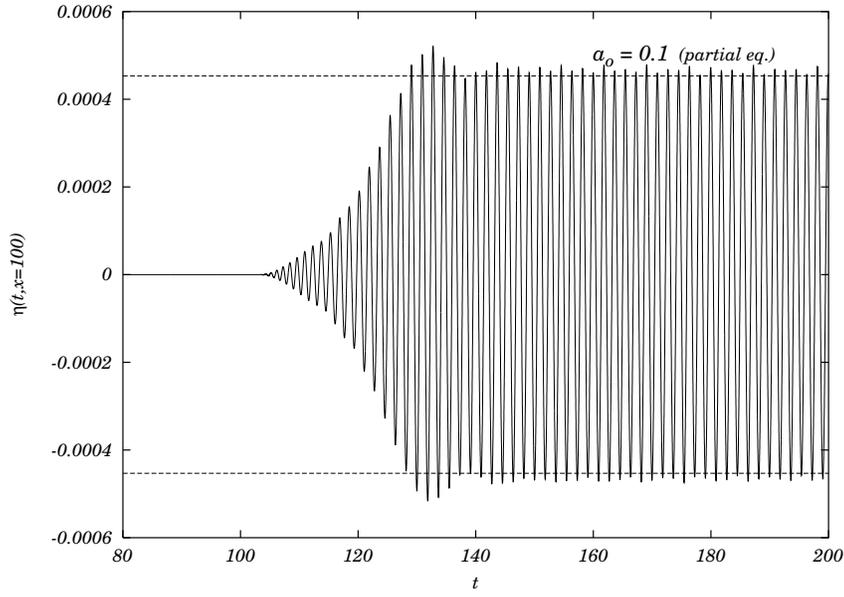}}}}
\end{picture}\\
\caption{Radiation computed from the full partial equation (1)}\label{calk2}
\end{center}
\end{figure}

\begin{figure}
\begin{center}
\begin{picture}(400,200)
\put(10,0){\rotate[r]{\resizebox{8cm}{!}{\includegraphics{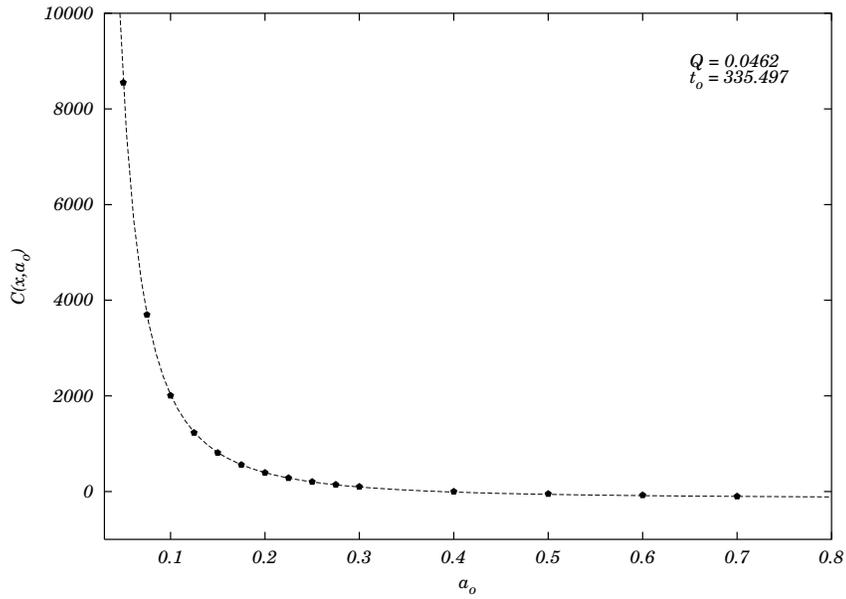}}}}
\end{picture}\\
\caption{$C(x,a_o)$ defined by eq. \rev{RXVIII} vs initial amplitude
of the discrete mode; numerical data and fitted function.}\label{fig2}
\end{center}
\end{figure}

\begin{figure}
\begin{center}
\begin{picture}(400,200)
\put(10,0){\rotate[r]{\resizebox{8cm}{!}{\includegraphics{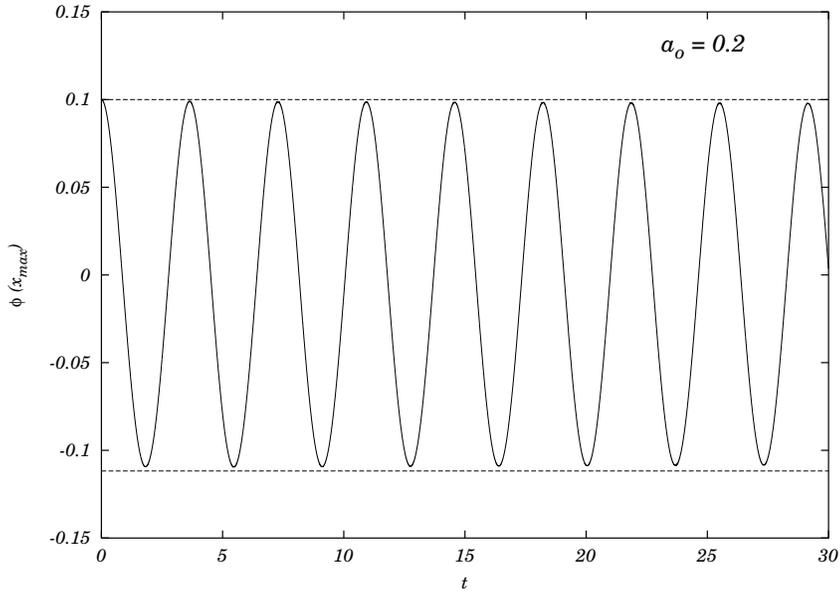}}}}
\end{picture}\\
\caption{Anharmonical oscillation of the discrete mode.
Dashed lines are calculated theoretically.}\label{anh} \end{center}
\end{figure}

\begin{figure}
\begin{center}
\begin{picture}(400,200)
\put(10,0){\rotate[r]{\resizebox{8cm}{!}{\includegraphics{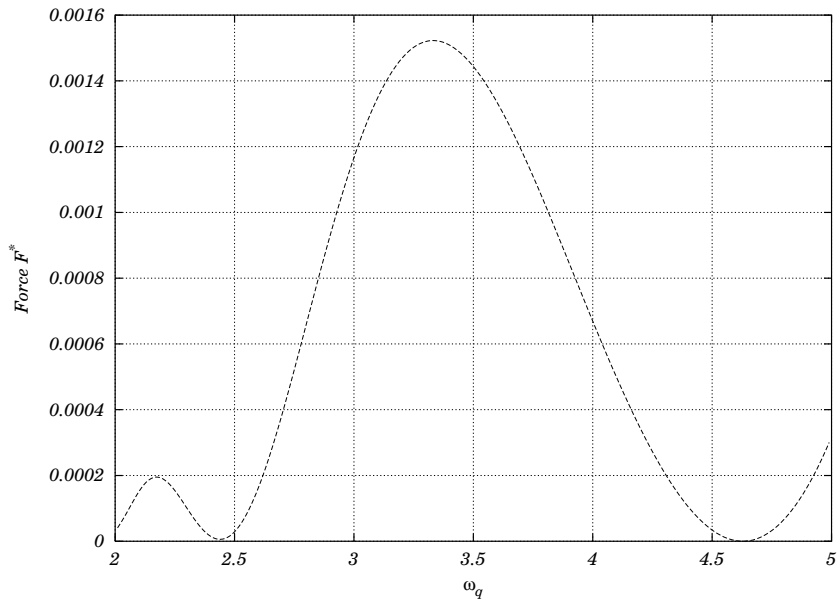}}}}
\end{picture}\\
\caption{Pulling force of the radiation vs frequency}\label{force} \end{center}
\end{figure}

\begin{figure}
\begin{center}
\begin{picture}(400,200)
\put(10,0){\rotate[r]{\resizebox{8cm}{!}{\includegraphics{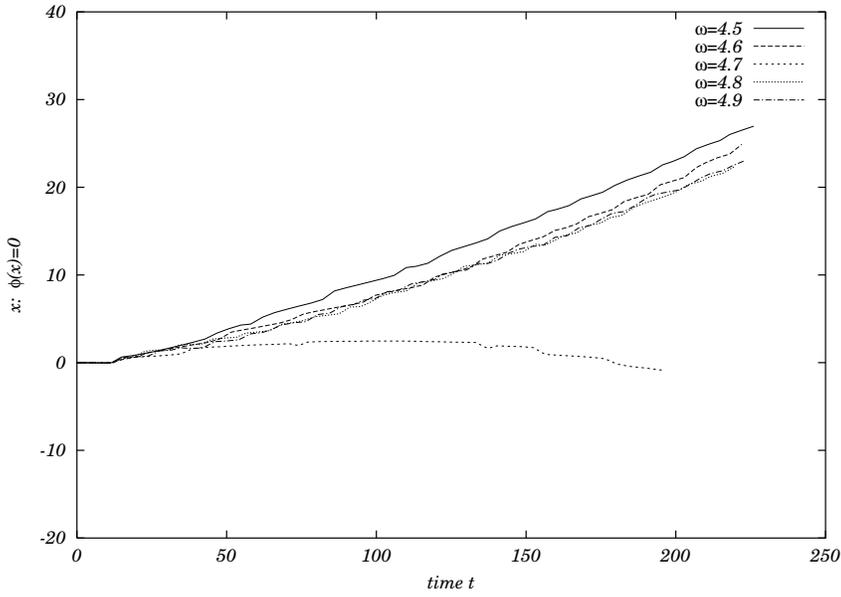}}}}
\end{picture}\\
\caption{Zeros of the field for different frequencies of the source wave and for the amplitude $a=0.2$}\label{wstecz1} \end{center}
\end{figure}

\begin{figure}
\begin{center}
\begin{picture}(400,200)
\put(10,0){\rotate[r]{\resizebox{8cm}{!}{\includegraphics{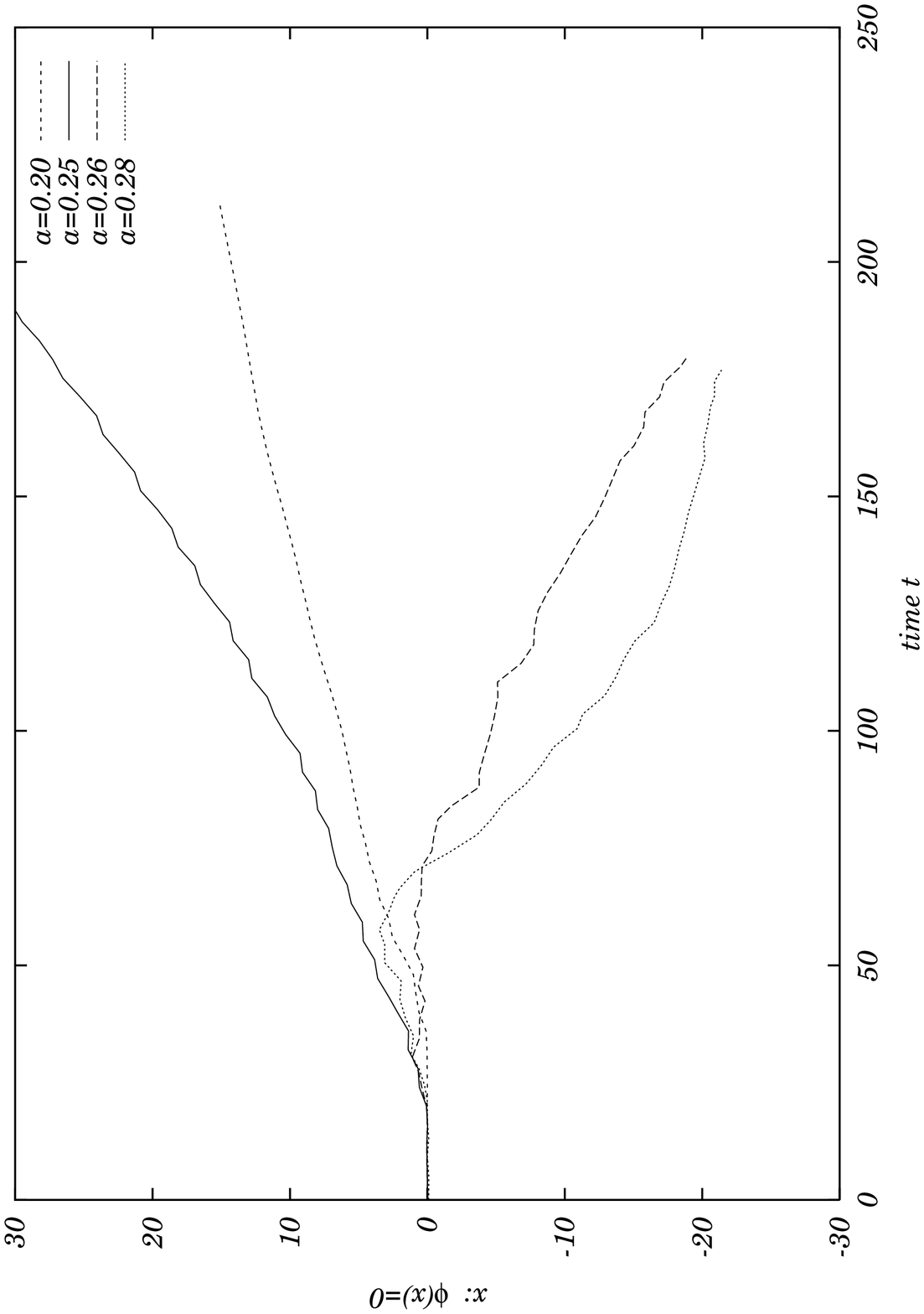}}}}
\end{picture}\\
\caption{Zeros of the field for $\omega=3.5$ and for different amplitudes}\label{wstecz2} \end{center}
\end{figure}


\begin{thebibliography}{25}
\bibitem{QCD} See eg. M.Baker, J.S. Ball, F. Zachariasen, Phys.Rep.
{\bf 209}, 73 (1991).
\bibitem{Cosmic} T.W.B. Kibble, J.Phys. {\bf A9}, 1387 (1979).
\bibitem{Saka} F.R. Bouchet, P. Peter, A. Riazuelo, M. Sakellariadou,
 Phys.Rev. {\bf D65}, 021301 (2002).
\bibitem{Kisel} V.G. Kiselev, Ya,M. Shnir, Phys. Rev. {\bf D57}, 5174
(1998).
\bibitem{Arodz1} H. Arod\'z, Nucl. Phys. {\bf B450}, 189 (1995).
\bibitem{Arodz2} H. Arod\'z, Acta Phys.Polon. {\bf B29}, 3725-3737
(1998).
\bibitem{Manton} N.S. Manton, H. Merabet, Nonlinearity {\bf 12}, 851
(1997).
\bibitem{Pelka} R. Pe\l ka, Acta Phys.Polon. {\bf B28}, 1981 (1997).
\bibitem{Slus} M. \'Slusarczyk, Acta Phys.Polon. {\bf B31}, 617-635
(2000).
\bibitem{Sengupta} R. Rajarshi, S. Sengupta, Phys. Rev. {\bf D65},
063521 (2002).
\bibitem{Habib}F.J. Alexander, S. Habib, A. Kovner, Phys. Rev. {\bf  E48},
4284-4296 (1993).

\end{thebibliography}
\end{document}